\documentclass{emulateapj}
\usepackage{natbib}
\usepackage{amsmath}
\usepackage{graphicx}
\usepackage{epsfig}

\def\Dwa{$\,$\uppercase\expandafter{\romannumeral5}$\,$}

\def\sless{\lower2pt\hbox{$\buildrel {\scriptstyle <}
   \over {\scriptstyle\sim}$}}

\def\sgreat{\lower2pt\hbox{$\buildrel {\scriptstyle >}
   \over {\scriptstyle\sim}$}}

\null\voffset=-2.0pc  

\begin{document}

\slugcomment{Received April 13, 2012; Accepted September 5, 2012}

\title{An Investigation into the Character of Pre-Explosion Core-Collapse Supernova Shock Motion}

\author{Adam Burrows\altaffilmark{1}, Joshua C. Dolence\altaffilmark{1}, Jeremiah W. Murphy\altaffilmark{1}}  
\altaffiltext{1}{Department of Astrophysical Sciences, Princeton University, 
Princeton, NJ 08544 USA; burrows,jdolence,jmurphy@astro.princeton.edu}

\begin{abstract}
We investigate the structure of the stalled
supernova shock in both 2D and 3D and explore the
differences in the effects of neutrino heating and the 
standing accretion shock instability (SASI). We find that
early on the amplitude of the dipolar mode of the shock 
is factors of $\sim$2$-$3 smaller in 3D than in 2D.
However, later in both 3D and 2D the monopole and dipole modes 
start to grow until explosion. Whereas in 2D the $(l,m) = (1,0)$
mode changes sign quasi-periodically, producing the ``up-and-down" motion
always seen in modern 2D simulations, in 3D this almost never
happens. Rather, in 3D when the dipolar mode starts to grow, it
grows in magnitude and wanders stochastically in direction until
settling before explosion to a particular patch of solid
angle. 
%
%
Furthermore, in 2D we find that the amplitude of the dipolar
shock deformation separates into two classes. For the first, 
identified with the SASI and for a wide range of ``low" neutrino
luminosities, this amplitude remains small and roughly constant.
For the other, identified with higher luminosities and neutrino-driven
convection, the dipolar amplitude grows sharply. Importantly, 
it is only for this higher luminosity class that we 
see neutrino-driven explosions within $\sim$1 second of bounce. 
Moreover, for the ``low" luminosity runs (including zero), the power 
spectra of these dipolar oscillations peak in the 30-50 Hz range 
associated with advection timescales, while for the high-luminosity 
runs the power spectra at lower frequencies are significantly more prominent.
We associate this enhanced power at lower frequencies with slower 
convective effects and the secular growth of the dipolar shock amplitude. 
Though our study involves a simplified, parametrized approach,
on the basis of it we hypothesize that neutrino-driven buoyant convection 
should almost always dominate the SASI when the supernova explosion is 
neutrino-driven.
%
\end{abstract}


\section{Introduction}
\label{intro}

Core-collapse supernova theory has evolved to the point that most researchers
now acknowledge a central role for aspherical, multi-dimensional hydrodynamics
in the mechanism of most such explosions.  The spherical (one-dimensional)
simulations explode only for the lowest-mass progenitors, and even then this
has been demonstrated only for one realization (the 8.8-M$_{\odot}$ model of Nomoto \& Hashimoto 1988) which
exploded with a tepid energy near $\sim$10$^{50}$ ergs (Kitaura et al. 2006; Burrows et al. 2007a). 
The pedigree for the recognition of the importance of instabilities extends into the 1970's, but the modern 
view emerged in the 1990s (Burrows \& Fryxell 1992; Herant, Benz, \& Colgate 1992; 
Herant et al. 1994; Burrows, Hayes, \& Fryxell 1995; Janka \& M\"uller 1996). 
Interpretations for the role of instabilities and turbulence are varied, but
include longer dwell times in the gain/heating region (Bethe \& Wilson 1985)
in multi-dimensional flows (Thompson, Quataert, \& Burrows 2005; Murphy \& Burrows 2008;
Janka 2001), altered cooling rates in the cooling region interior to the gain region
(Pejcha \& Thompson 2011), and turbulent fluxes and stresses (Murphy \& Meakin 2011).  It has also been claimed 
that large-amplitude dipolar ($l = 1$) oscillations of the stalled shock due to the 
standing accretion shock instability (SASI; Blondin, Mezzacappa, \& DeMarino 2003) 
facilitate explosion by (among other things) extending the gain volume and moving the shock to 
slightly larger stall radii. Such $l = 1$ modes are much in evidence in 2D, 
axisymmetric simulations, and the theory of this shock instability has been
well characterized and explored (Foglizzo 2002,2009; Foglizzo, Scheck, \& Janka 2006; Ohnishi et al. 2006;
Foglizzo et al. 2007; Scheck et al. 2008; Yamasaki \& Foglizzo 2008; Fern\'andez \& Thompson 2009a; 
Foglizzo et al. 2011,2012; Guilet \& Foglizzo 2012).  What has emerged is a model for the SASI involving 
a vortical-acoustic feedback cycle between the shock and the dense core.
In many relevant analyses, it is supposed that the amplitudes of the $l = 1$ modes
prevail and that such dipolar oscillations are a key to the relevance of the SASI 
in the core-collapse explosion context.

Importantly, if present, the SASI has been inherent in every multi-dimensional simulation of the
last $\sim$20 years, though some suppressed the $l = 1$ mode by their choice of
angular computational domain (Burrows, Hayes, \& Fryxell 1995; Buras et al. 2006).
Its identification and subsequent characterization,
while interesting in the context of our understanding
of supernova hydrodynamics, has not fundamentally
changed its potential role in facilitating neutrino-driven explosions.
%
%
What is new is its identification and the suggestion that
this shock instability, not neutrino-heating-driven buoyant convection,
is the important, perhaps central, multi-dimensional piece of the explosion 
puzzle.  Left implied is that, though neutrino heating might be crucial to explosion, 
the SASI generates the multi-dimensional turbulence that boosts 
neutrino heating's effect and accounts for the need to go to multi-D.  
However, the linear analyses of the instability, and the original 
hydrodynamic papers on the subject, neglected neutrinos 
all together.  In addition, the early hydro calculations highlighting or 
diagnosing the SASI were all done with simple $\gamma$-law equations of state 
(EOSs). Blondin, Mezzacappa, \& DeMarino (2003), using such $\gamma$-laws,
noted that the amplitudes of the SASI were stiffly increasing functions of 
$\gamma$.  Those authors originally suggested (but see Blondin \& Mezzacappa 2006) 
that the increasing transverse kinetic energy they found in their 
simulations might lead to explosion (without neutrinos).  However, 
they went on to show that when $\gamma$ was low (closer to the physical realm), 
the amplitudes of the standing shock instability were more tame.  

Despite this, though introduced as a shock instability with simple equations of state and no neutrinos,
there nevertheless emerged a tendency in many quarters to conflate the SASI with 
the turbulence seen in the more detailed multi-D simulations that included neutrino 
heating and the associated buoyant convection (Burrows et al. 2006,2007b; Ohnishi et al. 2006;
Bruenn et al. 2007,2010; Marek \& Janka 2009; Janka et al. 2007; Iwakami et~al. 2008; M\"uller, Janka, \& Marek 
2012). We conclude in this paper that such a tendency may have been unfounded.  
What is more, complete multi-D calculations done with multi-group neutrino transport have all 
been done in 2D, with axial symmetry. Such a symmetry suppresses non-axial flow 
patterns and instabilities, requiring, for example, the $l = 1$ mode to be an 
$m = 0$ mode as well. Suppressing the channeling of the free energy available
to non-zero $m$ modes and higher-order $l$ modes may exaggerate ``sloshing" along the axes and 
the magnitude of the $(l,m) = (1,0)$ mode\footnote{Moreover, those calculations done 
using multiple ray-by-ray radial transport in the 2D context may be artificially 
pumping up the $(l,m) = (1,0)$ mode by syncing large angular variations in 
the neutrino heating rates with the associated dipolar motions (Marek \& Janka 2009; 
M\"uller, Janka, \& Marek 2012; Bruenn et al. 2007,2010). Ott et al. 
(2008), Brandt et al. (2011), and Walder et al. (2005) have shown that more realistic angular 
distributions of the radiation fields and the concomitent neutrino heating profiles 
have much less angular variation than the associated turbulent and oscillating 
matter distributions.}.  We find this to be the case as well.  Therefore, in as much as it is associated 
with the vigor of the $(l,m) = (1,0)$ mode, the SASI's putative role in the 
neutrino mechanism may be exaggerated.

In fact, neutrino-driven buoyant convection may be the primary agency responsible 
for the turbulent motions behind the shock and the increased viability of explosion 
identified with multi-D post-shock turbulence (Murphy \& Meakin 2011), with the SASI subdominant.  This
was the conclusion of Fern\'andez \& Thompson (2009a) and had been the working 
hypothesis of supernova modelers since the 1990's.  A fraction ($\sim$4\% - 10\%) 
of the prodigious neutrino luminosities from the inner core is absorbed in 
the gain region. Such heating ``from below" and advection through the gain region
generate and maintain a negative entropy gradient that can
drive vigorous buoyant convection behind the stalled shock
and facilitate explosion through a variety of processes
still to be adequately characterized.
While several works have emphasized the potential role of
advection in stabilizing this region under some conditions
(Foglizzo et al. 2006; Yamasaki \& Yamada 2007),
Scheck et al. (2008) point out that these arguments are directly
applicable only in linear theory; nonlinear perturbations which
obtain in real supernova cores can lead to buoyant
convection even when linear theory suggests the region should
be stable. 

In this paper, we don't address all these issues. Rather, we focus specifically on comparing 
the motions of the shock 1) in two and three dimensions and 2) as a function of driving neutrino 
luminosity (including zero luminosity).  
%
%
What we find suggests that a reappraisal of the potential role of the SASI, 
vis \`a vis neutrino-driven convection in the core-collapse supernova context, is in order.
In part, this paper serves to inaugurate such a reappraisal.

In \S\ref{method}, we summarize our computational philosophy, emphasizing that 
we are not here employing detailed multi-dimensional, multi-group radiation hydrodynamics, 
but a more simplified approach to the neutrino sector.  In \S\ref{spherical}, we 
explore the spherical harmonic decomposition of the shock structure.
and investigate the stochastic motion of the dipole direction in 3D.  
We follow this in \S\ref{without} with a systematic study of the motion 
of the shock surface in 2D, with and without neutrinos and for a variety of driving luminosities, and conclude in 
\S\ref{conclusions} with a summary of basic findings.

\section{Calculations and Methodology}
\label{method}

We utilize the code CASTRO for these 3D and 2D calculations and comparisons 
of the collapse, bounce, and post-bounce evolution of the core of massive stars.
As described extensively in Almgren et al. (2010), 
CASTRO is a second-order, Eulerian, unsplit, compressible, Godunov radiation-hydrodynamics 
code that uses the piecewise-parabolic method. It has a 
multi-component advection scheme and uses adaptive mesh refinement (AMR).
Simulations can be performed in 1D (Cartesian, cylindrical, or spherical),
2D (Cartesian or cylindrical), and 3D (Cartesian). The AMR uses
block-structured hierarchical grids and each grid patch in 2D or 3D is
dynamically created or destroyed, logically structured, and rectangular.
The coarse grid provides Dirichlet data as boundary conditions for the
fine grids, the code subcycles in time, and a sophisticated synchronization
algorithm has been implemented. The software infrastructure rests on the
BoxLib foundation which has functionality for serial, distributed, and
shared-memory parallel architectures. The AMR framework employs flow control,
memory management, and grid generation. 

For these exploratory calculations, we use the ``light-bulb" neutrino heating 
and cooling algorithm of Murphy \& Burrows (2008) and Nordhaus et al. (2010a).
Constant electron-neutrino and anti-electron-neutrino luminosities (set equal to one another) 
are introduced after bounce that heat the shocked region by super-allowed charged-current 
absorption on free nucleons and the matter cools by the corresponding inverse processes.
The luminosities ($L_{\nu_e}$) used for the 3D portion of this study are $2.1\times 10^{52}$ ergs s$^{-1}$,
$2.2\times 10^{52}$ ergs s$^{-1}$, and $2.3\times 10^{52}$ ergs s$^{-1}$, and the neutrino
spectrum is that of a 4.0-MeV black body with zero chemical potential.
The deleptonization scheme of Liebend{\"o}rfer (2005) is used to evolve Y$_e$ (electron fraction)
exterior to the core. In the core, Y$_e$ is frozen after lepton trapping. While this 
methodology falls far short of real transport, it nevertheless enables one to witness 
shock stalling, post-bounce neutrino heating, neutrino-driven buoyant 
convection, and explosion phases in the context of state-of-the-art 
hydrodynamics.  We use 6 levels of refinement, with factors of two in spatial
zone scale between them and pre-zone the grid to reflect as best as we can the 
spherical nature of collapse flow.  In 3D, we use Cartesian coordinates and in 2D we use 
cylindrical ($r$ and $z$) coordinates. The entire domain from the center out to the 5000-kilometer
radius of an inscribed sphere is carried in the calculation and spherical, monopolar 
gravity is used.  The effective angular resolution near the shock (while interior
to a radius of $\sim$400 km) is 0.3$^{\circ}$ $-$ 0.6$^{\circ}$.
   
Since the Nordhaus et al. (2010a) work, a new version of the monopolar gravity routine in CASTRO that better handles the
multi-resolution AMR grid, in both 2D and 3D, has been implemented. In Nordhaus et al. (2010a), inaccuracies in the integrated mass
exterior to a $\sim$200 km radius, as well as corresponding differences in this quantity between 2D and 3D runs, were possible.  These
have been resolved by employing a more accurate algorithm and a much more spherical starting grid.  We are also using 2:1 (as opposed to 4:1)
jumps in resolution at refinement boundaries.  Though there are quantitative differences due to this more accurate scheme, and the
better resolution, there are no qualitative differences in the conclusions. In particular, the findings concerning the differences in
the character of the hydrodynamics in 3D and 2D and in the 3D/2D offset of the critical luminosity/mass-accretion-rate explosion curve
survive largely intact, though the differences in the critical luminosities between 2D and 3D are smaller (Dolence et al. 2012, in
preparation).
 
The progenitor model is the 15-M$_{\odot}$ star from Woosley \& Weaver (1995) 
that is non-rotating and has solar-metallicity at ZAMS (zero-age-main-sequence).
The equation of state is in tabular form as a function of $T$ (temperature), 
$\rho$ (mass density), and Y$_e$ and incorporates the Shen et al. (1998ab) 
nuclear mean-field theory.  We assume nuclear statistical equilibrium (NSE) 
and calculate the abundances of a single representative heavy nucleus at the peak of the shifting 
binding energy curve, alpha particles, nucleons, photons, electrons, and positrons.

The purpose of this study is not to diagnose the neutrino mechanism of explosion
in all its particulars and in great detail, but 1) to contrast the character and 
multipolarity structure of the stalled shock surface in 2D and 3D and 2) to compare 
the relative roles of neutrino-driven convection and the SASI. 
Our simpler approach (though not simple), avoiding as it does the computational 
burden of full transport while providing a sophisticated hydrodynamics capability, 
is designed to clarify the qualitative differences between 2D and 3D. 
We find that there are many such differences, and now proceed 
to a discussion of some of the most interesting among them.

\section{Spherical Harmonic Decomposition of Shock Structure: 2D versus 3D}
\label{spherical}

Since spherical harmonics form a complete orthonormal basis set for a two-dimensional surface,
at any particular time one can decompose the closed surface of the shock ($R_s(\theta,\phi)$)
into spherical harmonic components with coefficients:
\begin{equation}
a_{lm} = \frac{(-1)^{|m|}}{\sqrt{4\pi(2l+1)}} \oint R_s(\theta,\phi) Y_l^m(\theta,\phi) d\Omega\, ,
\end{equation}
where we normalized such that $a_{00} = a_{0} =\langle R_s\rangle$, the average
shock radius, and $a_{11}$, $a_{1{-1}}$, and $a_{10}$ correspond to the average
Cartesian coordinates of the shock surface $\langle x_s\rangle$, $\langle y_s\rangle$,
and $\langle z_s\rangle$, respectively. The magnitude and principle axis of the
dipolar shock deformation are then trivially obtained from these Cartesian moments.
The orthonormal harmonic basis functions are 
\begin{equation}
Y_l^m(\theta,\phi) = \begin{cases} 
	\sqrt{2} N_l^m P_l^m(\cos\theta) \cos m\phi&		m>0\, ,\\
	N_l^0 P_l^0(\cos\theta) &				m=0\, ,\\
	\sqrt{2} N_l^{|m|} P_l^{|m|}(\cos\theta) \sin |m|\phi&	m<0\, ,
\end{cases}
\end{equation}
where
\begin{equation}
N_l^m = \sqrt{\frac{2l+1}{4\pi}\frac{(l-m)!}{(l+m)!}}\, ,
\end{equation}
$P_l^m(\cos\theta)$ are the associated Legendre polynomials, 
and $\theta$ and $\phi$ are the spherical coordinate angles.

%

Furthermore, we calculate an angular ``energy," $P(l)$, of the function $R_s(\theta,\phi)$ on the unit sphere as:
\begin{equation}
\frac{1}{4\pi}\int_{\Omega} R_s(\theta,\phi)^2\ d\Omega
= \sum_{l = 0}^{\infty}\ P(l)\, ,
\end{equation}
where 
\begin{equation}
P(l) = \sum_{m = -l}^{l} a_{l{m}}^2\, .
\label{pl}
\end{equation}

Figure \ref{fig1} depicts snapshots for various times after bounce of 
slices colored by entropy of our 2D and 3D simulations using 
$\rm{L_{\nu_e}} = 2.1\times 10^{52}$ ergs s$^{-1}$. We see that the axial symmetry and 
large-amplitude dipolar excursions, so much in evidence in 2D, are missing in 3D. 
Moreover, Figure \ref{fig1} clearly indicates that there is much more small-scale 
structure in 3D than in 2D.  All in all, Figure \ref{fig1} suggests that there 
are numerous qualitative differences between runs that differ only in dimension.

Figure \ref{fig2} depicts the temporal evolution (after bounce) of the magnitude of the $a_{l{m}}$ coefficients, normalized
to the monopole term, of the shock wave surface for both 2D and 3D simulations.  Various representative driving luminosities
are employed to show the associated trend(s).  For the 3D runs, we show the square root of the sum of the squares 
of the $a_{l{m}}$ coefficients for a given value of $l$ at the different associated azimuthal quantum numbers
\footnote{Note that this means the $(l,m) = (1,0)$ mode itself is smaller than this composite number.}, $m$,
also normalized to the corresponding monopole term. 
This quantity is equal to $\sqrt{P(l)}$.  

As Figures \ref{fig2}, \ref{fig3}, and \ref{fig4} indicate, the larger axial excursions in $|a_{1}|/a_{0}$ seen in 
2D are almost absent in 3D (Iwakami et al. 2008). For the lower luminosity ($2.1\times 10^{52}$
ergs s$^{-1}$) model (for which here explosion occurs later than shown) and for all models
during {\it early} phases (first few hundred milliseconds - see Figure \ref{fig2}), the time-averaged 
3D dipolar amplitudes are generally factors of $\sim$2$-$3 lower than those in 2D.  
Figure \ref{fig3} depicts on a Mollweide projection the instantaneous direction (dot position) and magnitude
(dot size) of the dipolar mode in 3D as a function of time (dot color), for two
representative models.  Importantly, Figure \ref{fig3} shows that in 3D the direction 
of the dipole jumps around (Kotake et al. 2009), while its magnitude secularly evolves. In the march to 
explosion, it eventually settles on a particular direction (or small patch of solid angle), 
increasing in magnitude but not changing direction (or sign).  Figure \ref{fig4} depicts the individual 
$a_{lm}$ coefficients (not their absolute values), for the dipoles in 3D and 2D, and tells the same story.  
Though the dipole's direction jumps about stochastically at early times, as the march towards explosion 
commences (which could be many hundreds of milliseconds prior to explosion)
the dipole's general direction settles.  In particular, at these later times, the dipolar component
can cease changing sign long before the corresponding 2D model.  

Consistently, the $l=1$ coefficients vary more wildly in 2D at all times than the 
corresponding summed values in 3D. The behavior in 3D is more smooth.  
Axial sloshing along one direction has been a signal characteristic of 2D simulations, and a vigorous,
direction-changing $(l,m) = (1,0)$ mode has been identified in the past as an aid to
explosion (e.g., Marek \& Janka 2009; Hanke et al. 2012).  

Nevertheless, as Figures \ref{fig2} and \ref{fig4} indicate, we do see an 
increase in the amplitudes of the $l=1$ modes in both 2D and 3D simulations 
as explosion is approached and post-shock convection in and around
the gain region becomes more vigorous. Moreover, the growth in the $l=1$ seems to 
precede the growth in the $l=0$ (average radius) mode (see Figure \ref{fig10} below).
In 3D, we obtain characteristic growth times in $|a_{1}|/a_{0}$ with 
these constant driving luminosities of $\sim$50$-$300 milliseconds.  
In 3D, the near lack of a direction change in the dipolar component at 
late times, accompanied by a steadily increasing amplitude, 
may be more conducive to explosion than the frequent direction change seen in 2D 
(Dolence et al. 2012, in preparation).  

In 3D, as Figure \ref{fig2} suggests, we find that the 
``energy" in the $l = 1$ mode is spread over the various 
$m$ submodes, and is not exclusively in the $m=0$ component.
This was a conclusion of Blondin \& Mezzacappa (2007) as well (see 
also Yamasaki \& Foglizzo 2008, Fern\'andez 2010, Iwakami et al. 2008, 
Kotake et al. 2009, and Foglizzo et al. 2012), though Blondin \& Mezzacappa (2007) 
were most interested in induced differential rotation\footnote{which Rantsiou et al. (2011) 
suggest is not large when integrated over the entire proto-neutron star} through the unbalanced growth of $m=1,-1$ modes.
In addition, we find that in 3D a larger fraction of the total modal energy is partitioned 
into higher-$l$ modes (with smaller angular scales) than in 2D. The behavior 
of $|a_{3}|/a_{0}$ in Figure \ref{fig2} shows this (as do many of the other higher-order $l$ modes, not shown). 
The quadrupolar modes ($|a_{2}|/a_{0}$), however, seem to reside near the juncture between 
these two behaviors. In 2D convection, it is known there is an enhancement of 
power on larger scales (Boffetta \& Ecke 2012), whereas in 3D there is an energy cascade 
to smaller scales (Itoh \& Itoh 1998). This results in the tilting towards 
larger $l$s for 3D convection and seems to explain what we see 
(see also Hanke et al. 2012). The net result is a redistribution of power to 
smaller structures and larger $l$s in 3D than in 2D and the 
untethering of the ``up-and-down" $m=0$ dipolar mode.   

In the context of the differences between
2D and 3D turbulence, it has been
shown that as long as the Mach number
is not too large many ideas derived from incompressible
turbulence are relevant for compressible turbulence.
Finally, both numerical and experimental studies lend
credence to the idea of an inverse energy cascade in
compressible 2D turbulence (e.g., Dahlburg et al. 1990; 
Antar 2003; Boffetta \& Musacchio 2012). 

In summary, we find that, for these non-rotating models, the $(l,m) = (1,0)$/sign-changing axial mode, 
so prominent in 2D as vigorous up-and-down sloshing along the symmetry axis, 
is not much in evidence in 3D simulations at any stage. A dipolar mode in a randomly 
established direction does grow without varying in direction at later times before and during explosion.
In addition, we see in 3D, for the higher luminosities and/or at later times near explosion,
a steady increase in amplitude with time for the lower-order modes 
(cf. Figures \ref{fig2} and \ref{fig4}). This behavior may be indicative 
of the character of neutrino-driven explosions, but whether this is generically the case 
in realistic 3D neutrino-driven supernova explosions remains to be seen (Takiwaki et~al. 2011).

\section{Shock Behavior: The Role of Neutrino Luminosity}
\label{without}

Ostensibly, the effects of buoyancy-driven convection and the ``SASI" shock instability are intimately entangled 
in Nature and it might be difficult to separate out their respective roles in supernova dynamics. 
Moreover, the early ``SASI" studies had been analytically diagnosed and explored  
without neutrino heating (Blondin, Mezzacappa, \& DeMarino 2003; Blondin \& Mezzacappa 2006; 
Foglizzo 2002,2009; Foglizzo et al. 2007; Yamasaki \& {Foglizzo 2008).
This may have led some to confuse the ``SASI" with the convection and instabilities
that are seen in the multi-dimensional simulations that include neutrino transport and 
transfer.  One, therefore, would like to separate out the two instabilities, or at least to 
isolate one from the other in the same numerical context, that nevertheless incorporates
a reasonable progenitor and equation of state.

Since with neutrinos on (and with a stalled shock) the SASI can't easily be distinguished or isolated, 
we first simulate two 2D models, one with and one without neutrino heating.  In order to
ensure that the shock stalls at a reasonable radius (100$-$200 km) and that the accumulation
of accreted matter does not push the shock inexorably outwards, for the model without 
neutrino heating we have left on electron capture (via the Liebend\"orfer prescription) 
and included minimal neutrino cooling at 10\% of its normal rate\footnote{When we perform 
the simulation with no cooling whatsoever, the average shock radius grows from $\sim$200 km to $\sim$500 km,
undermining the attempt to maintain the shock radius within the canonical pre-explosion range.}.  
Also for the ``no-neutrino-heating" model, during the first 50 milliseconds after bounce 
we first allow the shock to stall and settle with neutrino heating on, 
and then turn off the heating and cut the cooling as indicated.
Otherwise, for both runs we use the same progenitor and nuclear equation of state. 
By this procedure, we ensure that the pre-shock speeds and the mean stalled shock 
radii are near those in the normal core-collapse context, while completely shutting off
neutrino heating everywhere, in particular in the gain region, and thereby neutrino-driven 
convection.  The goal is to compare the vigor of shock motions in these two 
cases.

Figure \ref{fig5} portrays representative entropy maps during our $\sim$1-second comparison
test of 2D models with and without neutrino heating.  This figure clearly shows 
that this ``no-neutrino-heating" model is more quiescent and spherical 
than the model with neutrino heating. The SASI evolution seems 
inadequate to explain the vigor of the turbulence found in full
simulations of core collapse.  Hence, this comparison suggests 
that neutrino-driving is the dominant agency of turbulent dynamics (Murphy et al. 2012), with the 
shock instability subdominant. This conclusion is in keeping with
the central role expected for neutrino heating in powering neutrino-driven 
explosions.

Figure \ref{fig6} shows the evolution of the $l=(1,2,3)$ harmonic 
coefficients of the shock surface in the two cases. The amplitudes of the $l = 1$ dipolar 
mode when neutrino heating is absent are factors of $\sim$3$-$4 smaller than when neutrino heating 
is turned on.  It is possible that photodisocciation at the shock (and the associated effect 
on the pressure), absent from the analytic studies, are playing a role here in muting 
the SASI amplitudes (Fern\'andez \& Thompson 2009b). Importantly, however, it is only when we turn on 
neutrino heating that convection becomes vigorous, the specific entropy increases significantly, 
and the amplitudes of the shock oscillation and turbulent convection in the gain region 
are significant. 

However, though Figures \ref{fig5} and \ref{fig6} clearly indicate
qualitative differences, our choice of parameters for these models 
(in particular the reduced cooling model) may have biased our conclusions.  
Therefore, we follow this binary test with a series of ten similar 
calculations for different driving luminosities, $L_{\nu_e}$, 
but with full cooling turned on.  The corresponding $L_{\nu_e}$s are zero,
0.5, 1.0, 1.5, 2.0, 2.1, 2.2, 2.23, 2.3, and 2.33 $\times 10^{52}$ ergs s$^{-1}$.

Figure \ref{fig7} depicts the root-mean-square (rms) amplitude 
$\sqrt{\langle a_1^2\rangle}$ of the dipolar shock deformation as a function of
neutrino luminosity for all ten models. Note that here we do not normalize to $a_{0}$,
though the rms amplitude of the ratio $a_{1}/a_{0}$ has the same trends as seen in Figure \ref{fig7}.
Below $L_{\nu_e}\sim2\times10^{52}\,{\rm erg\,s}^{-1}$, the amplitudes remain small and
roughly constant even though the average shock radii change by a factor $\sim$two.  For higher luminosities
(shaded region), the amplitudes grow steeply; we associate this region with the onset of vigorous neutrino-driven
convection. This is consistent with the dramatic increase in low-frequency power.
Figure \ref{fig8} portrays power spectra of the dipolar shock deformation $a_1(t)$ 
for models with different neutrino luminosities (in units of $10^{52}\,{\rm erg\,s}^{-1}$).  
Also shown is the corresponding quantity for $a_1(t)$ normalized by $a_0(t)$.
Models with $L_{\nu_e}\ge2\times10^{52}\,{\rm erg\,s}^{-1}$ show
$\sim${\it 100 times} more low-frequency power than models with lower luminosities.  
Such slower motions are most sensibly associated with the slower convective plumes
of buoyancy-driven convection. The excess low-frequency power is 
\emph{not} associated with explosion; models with $L_{\nu_e}\le2.1\times10^{52}\,{\rm
erg\,s}^{-1}$ do not explode within the simulated time and only times before explosion are included in this
analysis for more luminous models. 

A parameter that Foglizzo et al. (2006) and Scheck et al. (2008) both suggest
using to diagnose the potential for advection through the gain 
region to stabilize buoyancy-driven convection is $\chi$, an average of 
the ratio of the advection time to the Brunt overturn timescale in this region:
\begin{equation}
\chi = \int^{R_s}_{R_{gain}} \frac{\omega}{|v|} dr\, ,
\label{brunt}
\end{equation}  
where $\omega$ is the Brunt-V\"ais\"al\"a frequency, $r$ is the radius, 
and $|v|$ is the absolute value of the radial velocity. In calculating eq. (\ref{brunt}),
we use spherically-averaged values of the density, pressure, $|v|$, and adiabatic $\gamma$.
While Foglizzo et al. (2006) and Scheck et al. (2008) emphasized the potential 
role of advection in stabilizing this region under some conditions,
Scheck et al. (2008) point out that these arguments are directly
applicable only in linear theory; nonlinear perturbations which
actually obtain in real supernova cores can lead to buoyant
convection even when linear theory suggests the region should
be stable.  Also, according to Scheck et al. (2008), the idea of a critical
$\chi$ for convective stability doesn't apply when there are perturbations
of O(1\%).  Such perturbations may be found in the pre-collapse progenitors 
(Arnett \& Meakin 2011).  

Nevertheless, in Figure \ref{fig9} we plot the evolution 
of a smoothed value of $\chi$ for most of our 2D luminosity study runs, 
from ``low" to ``high" values. We see that, for a given $L_{\nu_e}$, the value of $\chi$ 
can start low, but quickly settles to a constant value that increases 
monotonically with driving neutrino luminosity.  Moreover, if we use the
critical $\chi$ of 3 suggested in Foglizzo et al. (2006), this value
can be seen (roughly) to separate our two classes.
In fact, if we were to strictly adhere to the $\chi=3$ discriminant
we might conclude that we have been a bit too conservative in Figure \ref{fig7}
in distinguishing our two classes and that the ``SASI-like" class might encompass 
a narrower range bounded from above near $L_{\nu_e} = 1.5\times10^{52}\,{\rm erg\,s}^{-1}$
instead of near $L_{\nu_e} = 2.0\times10^{52}\,{\rm erg\,s}^{-1}$. 
One of the reasons the value of $\chi$ is larger for the larger 
neutrino luminosities is simply that the shock radii and gain regions 
are larger for larger luminosities.  This alone might suggest that
an important condition for the success of the neutrino heating mechanism,
that the shock radii and gain regions are large, also translates into a high $\chi$.
By the logic of the use of $\chi$ to determine the importance of buoyancy-driven
convection, this might then separately suggest that viable neutrino-driven explosion 
models are also those for which the turbulence is buoyancy-driven.  
In any case, it is clear that both Figures \ref{fig8} and \ref{fig9}
suggest the bifurcation into the two luminosity classes we have 
described has merit. 

Therefore, we find that the amplitude of the dipolar
shock deformation remains small and roughly constant for a wide
range of low luminosities, but that it grows sharply at higher
luminosities.  The power spectra of the dipolar
shock oscillations separate into two classes. 
For the low luminosity runs, they peak in the 30$-$50 Hz range, comparable to the
timescale one might expect for an advective-acoustic cycle. 
For the high-luminosity runs, they are dominated
by low frequencies, where they have much more power than the
lower-luminosity models.  Taken together, these results suggest that
there is a SASI-dominated family (``low" luminosity models) and
a neutrino-driven convection-dominated family (``high" luminosity
models).  Importantly, however, we find that all the models that
explode within $\sim$1 second of bounce are well within the region of
parameter space where neutrino-driven convection dominates. We
conclude that, at least for this particular progenitor and in the
context of our simplified modeling, neutrino-driven convection
dominates whenever there is a neutrino-driven explosion.


\section{Discussion and Conclusions}
\label{conclusions}

In this paper, we have investigated the differences between the character of the motions  
of the stalled supernova shock in both 2D and 3D.  In addition, we have attempted
to distinguish between the effects of buoyant convection driven by neutrino heating and the 
standing accretion shock instability (SASI).  The latter is a hydrodynamic instability
that has been implicated in the vigorous dipolar ($l = 1$) shock oscillations seen in 2D 
axisymmetric supernova simulations (e.g., Burrows et al. 2006,2007b; Buras et al. 2006; Mezzacappa et al. 2007; 
Bruenn et al. 2007,2010; Marek \& Janka 2009; M\"uller, Janka, \& Marek 2012), even those with neutrino transport and 
neutrino-matter coupling, and is thought by some to be important in the supernova explosion phenomenon. 
We find, however, that in 3D the $(l,m) = (1,0)$ dipolar sloshing mode, putatively a signature of the SASI, is 
subdominant and not a major feature of shock dynamics. In 3D, the free energy available to 
instability seems to be bled from the  $(l,m) = (1,0)$ mode into other modes, both 
the $(l,m) = (1,\pm{1})$ (Blondin \& Mezzacappa 2007; Iwakami et al. 2008; Kotake et al. 2009) 
and the higher-order $l$ modes, and perhaps, in part, into the $l = 0$ (shock radius) mode. 
In fact, as Figure \ref{fig10} shows, in 3D we frequently see a quasi-steady increase 
in $a_{0}$, the average shock radius, and in $|a_{1}|$, the magnitude of the (vector) dipole,
even hundreds of milliseconds before the explosion clearly commences.
Such a slow, inexorable, increase in both shock radius and a unidirectional 
dipolar component may be distinctive and important signatures of 3D 
supernova behavior in the neutrino-driven context.  

Also, early in the non-linear convective stage when the shock 
is stalled, the amplitudes of the $l = 1$ mode are factors of $\sim$2$-$3
smaller in 3D than in 2D.  In addition, many hundreds of milliseconds before explosion 
in both 3D and 2D the low-order modes (e.g., the $l = 0,1,2,3$ modes) often start to grow until explosion. However, 
whereas in 2D the $(l,m) = (1,0)$ mode changes sign quasi-periodically, producing the ``sloshing" and
``up-and-down" motion always seen in modern 2D supernova simulations (Burrows et al. 2006,2007b;
Buras et al. 2006; Bruenn et al. 2007,2010; Marek \& Janka 2009; Janka et al. 
2007; M\"uller, Janka, \& Marek 2012), in 3D such motion is not much in evidence. On 
the contrary, when the dipolar mode starts to grow, in 3D it grows slowly 
in magnitude and wobbles ``randomly" in direction, until from less than one hundred 
to a few hundred milliseconds before explosion it settles to a general direction.  
The result is qualitatitively different shock behavior and hydrodynamic development in 3D and  
little or no ``sloshing" motions, such as are seen in 2D. It remains to be determined, however, whether 
this qualitatively different behavior is a consequence of the assumptions of our simple setup,
or are generic features that will survive when full radiation/hydrodynamic simulations 
in 3D are finally, and credibly, performed.

In 2D, we find that the absolute amplitude of the dipolar
shock deformation remains small and roughly constant for a wide
range of ``low" luminosities (including zero), but that it grows 
sharply at higher luminosities.  The power spectra of the dipolar
shock oscillations separate these ``high" and ``low"
luminosity runs into two classes, with those for the ``low"
luminosity runs peaking in the 30$-$50 Hz range and those of 
the high-luminosity runs dominated by low frequencies, where they 
have $\sim$100 times more power. Taken together, these results suggest 
that there is a SASI-dominated family and a neutrino-driven,  
convection-dominated family. Since neutrino-driven explosions require 
ample neutrino heating, we think that the latter represents the more 
realistic branch. The SASI could be in evidence in the latter, but 
seems not determinative of the dynamics, nor as centrally important 
to the outcomes as the buoyant plumes and bubbles of neutrino-driven convection.  

We find that all the models that explode within $\sim$1 second of bounce are well 
within the region of parameter space where neutrino-driven convection dominates and 
emphasize that our apparently SASI-dominated runs never explode. It is possible
that after bounce the SASI could grow earlier than buoyancy-driven 
convection, but that its central role is to seed the 
neutrino-driven buoyancy instability, which would, according to our calculations 
and our identification of the ``high-luminosity" class with most real 
core-collapse supernovae, dominate thereafter. It is possible, though, 
that perturbations in the progenitor could play this role as well (Arnett \& Meakin 2011).

Therefore, on the basis of our simulations, though they were done 
with one particular progenitor and were performed with a simplified 
physical approach, we are led to hypothesize that the SASI instability 
is subdominant in supernova dynamics when the supernova explosion is
neutrino-driven.  However, full 3D numerical simulations, with well-crafted 
diagnostics, are needed to prove this hypothesis definitively.
%
 
Since many early theoretical studies of the SASI ignored neutrino physics,
in particular neutrino heating, and used a simple gamma-law 
equation of state, we surmise that this led some researchers performing 
full 2D supernova simulations (and others) to connect uncritically the vigorous dipolar 
modes seen invariably in 2D with the very real instability that is the SASI. Moreover, 
in 2D (unlike in 3D), the tendency for both the SASI and buoyant convection to shunt power into 
low-order modes and for the turbulent energy cascade to be inverse (Boffetta 
\& Musacchio 2012; Hanke et al. 2012) might have led to some confusion.  
%

%

Hanke et al. (2012) have suggested that 1D and 3D behaviors are quite similar,
become more so as resolution is improved, and that something like the vigorous 
dipolar motion seen in 2D might be necessary for explosions in the 3D of Nature. 
While we agree that 3D is different from 2D, we don't agree that it becomes 
more similar to 1D, and we don't agree that the sloshing seen in 2D is a 
necessary condition of explosion.  However, we note that our 
3D simulations, though performed with simplistic neutrino transfer, reveal 
a growing dipolar mode (along with growing higher-order and $l = 0$ modes) 
as explosion is approached. Importantly, we find that the dipole term does 
not oscillate quasi-periodically in an up-and-down axial motion, 
but grows in ``random" directions, as Figures \ref{fig2}, \ref{fig3}, and \ref{fig4}  
imply, until the approach to explosion settles on a direction. Also, its growth
tends to precede that of the $l=0$ mode. 
In addition, it is becoming clear that a dipolar mode is necessary 
to generate pulsar kicks (Scheck et al. 2004,2006; Wongwathanarat 
et al. 2010; Nordhaus et al. 2010b,2012). A growing dipolar mode, 
similar to what we find in 3D that nevertheless does not partake 
of the sloshing behavior identified in detailed 2D simulations to 
date, is consistent with this expectation.  

We have not included in this study models with some initial rotation. 
The presence of a rotation axis might set a direction that could be 
of relevance in the hydrodynamic partitioning of ``energy" into $l$ 
and $m$ modes and submodes.  Furthermore, our 3D and 2D studies did not
incorporate detailed neutrino transport.  Nevertheless, the qualitative differences we see 
between 2D and 3D hydrodynamic behavior and the subdominance of the SASI
in the necessary context of strong neutrino heating and the resulting buoyancy-driven 
convection would seem to be important conclusions of this investigation.

\acknowledgments
The authors acknowledge stimulating interactions with Jason Nordhaus, Christian Ott, 
Rodrigo Fern\'andez, John Blondin, Ann Almgren, John Bell, and Manou Rantsiou. 
A.B. acknowledges support from the Scientific Discovery through Advanced Computing (SciDAC) 
program of the DOE, under grant number DE-FG02-08ER41544, the NSF under the subaward 
no. ND201387 to the Joint Institute for Nuclear Astrophysics (JINA, NSF PHY-0822648),
and the NSF PetaApps program, under award OCI-0905046 via a subaward 
no. 44592 from Louisiana State University to Princeton University.
The authors would like to thank the members of the Center for Computational 
Sciences and Engineering (CCSE) at LBNL for their invaluable support for CASTRO.
The authors employed computational resources provided by the TIGRESS
high performance computer center at Princeton University, which is jointly supported by the Princeton
Institute for Computational Science and Engineering (PICSciE) and the Princeton University Office of
Information Technology; by the National Energy Research Scientific Computing Center
(NERSC), which is supported by the Office of Science of the US Department of
Energy under contract DE-AC03-76SF00098; and on the Kraken supercomputer,
hosted at NICS and provided by the National Science Foundation through
the TeraGrid Advanced Support Program under grant number TG-AST100001.


\newpage

\begin{figure}
\begin{center}
\includegraphics[height=.25\textheight]{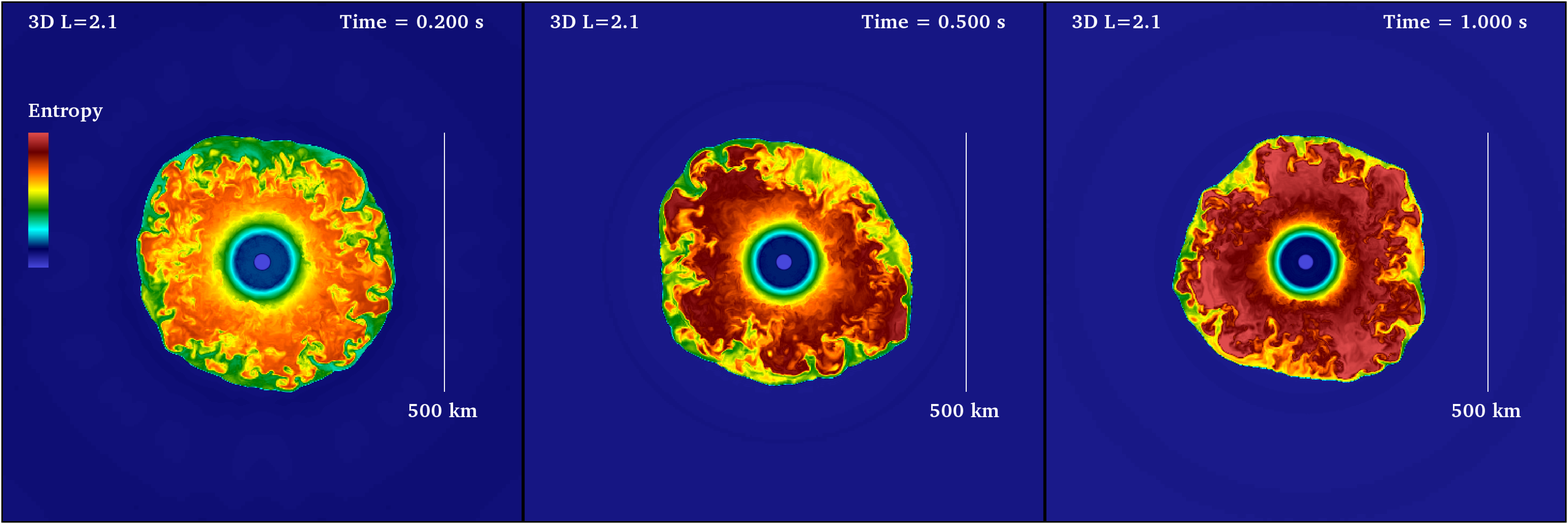}
\includegraphics[height=.25\textheight]{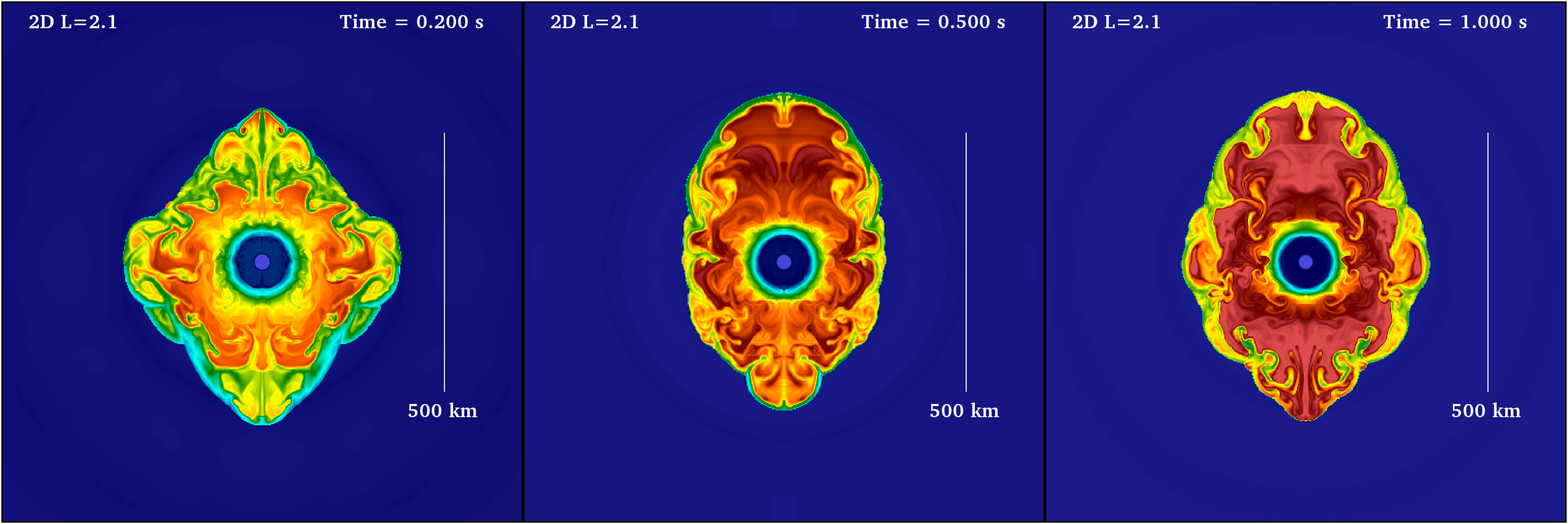}
\end{center}
\caption{A comparison of time slices of 3D (top) and 2D (bottom) runs of the entropy distributions after bounce
for a representative model for which $L_{\nu_{e}} = 2.1\times 10^{52}$ ergs s$^{-1}$.
The color scale ranges from 2.5 (blue) to 17 (red) $k_b/\textrm{baryon}$. The three 
different stills for each set (3D and 2D) are meant to depict characteristic
structures seen in each context during the evolution towards explosion.  As this figure
suggests, the large-scale dipolar features seen in 2D are absent in 3D.  In addition, there is
more small-scale structure in 3D.  Please refer to the 
discussions in the text for further clarifications.} 
\label{fig1}
\end{figure}

\clearpage

\begin{figure}
\includegraphics[height=.35\textheight]{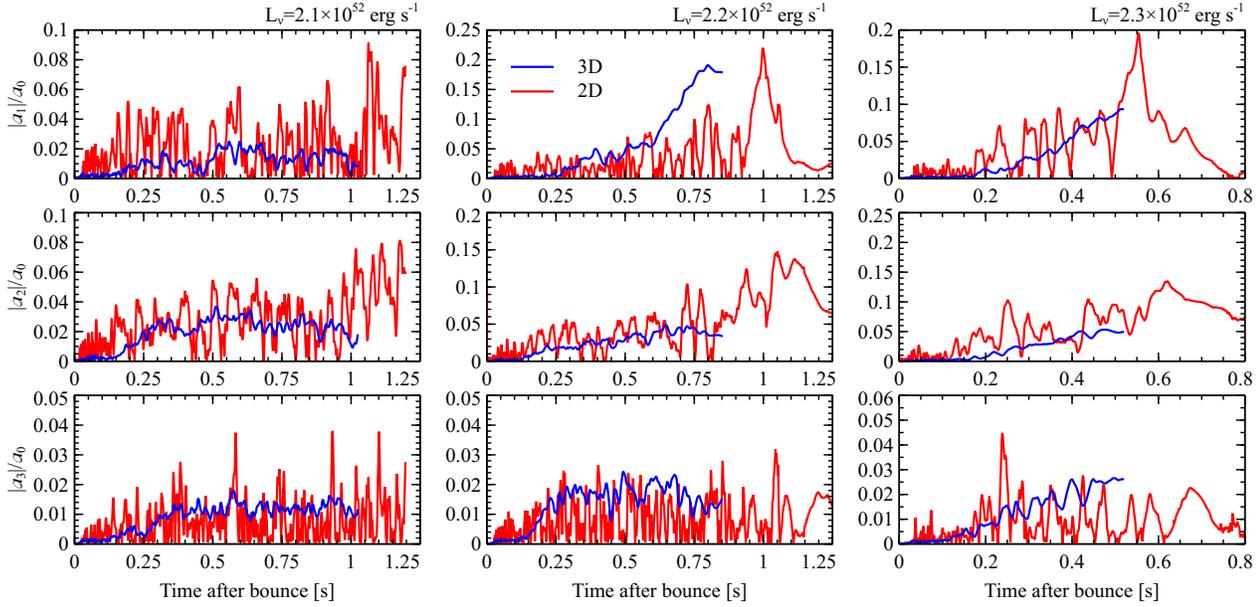}
\caption{The evolution of the amplitude of the harmonic coefficents 
of the shock wave surface in 2D (red) and 3D (blue),
as rendered by the square root of $P(l)$ (eq. \ref{pl}), 
for $l = 1,2,{\rm and}\ 3$ for three models with 
$L_{\nu_{e}} = (2.1\times 10^{52}, 2.2\times 10^{52},\ {\rm and}\ 2.3\times 10^{52}$) ergs s$^{-1}$.
The saturation amplitudes of the dipolar term for the lowest $L_{\nu_{e}}$ model (on left), 
for which explosion is not in evidence during the first second, are invariably lower in 3D. 
Note that the scales of the various panels are not the same.  
See text for a discussion.}
\label{fig2}
\end{figure}

\clearpage

\begin{figure}
\begin{center}
\includegraphics[height=0.45\textheight]{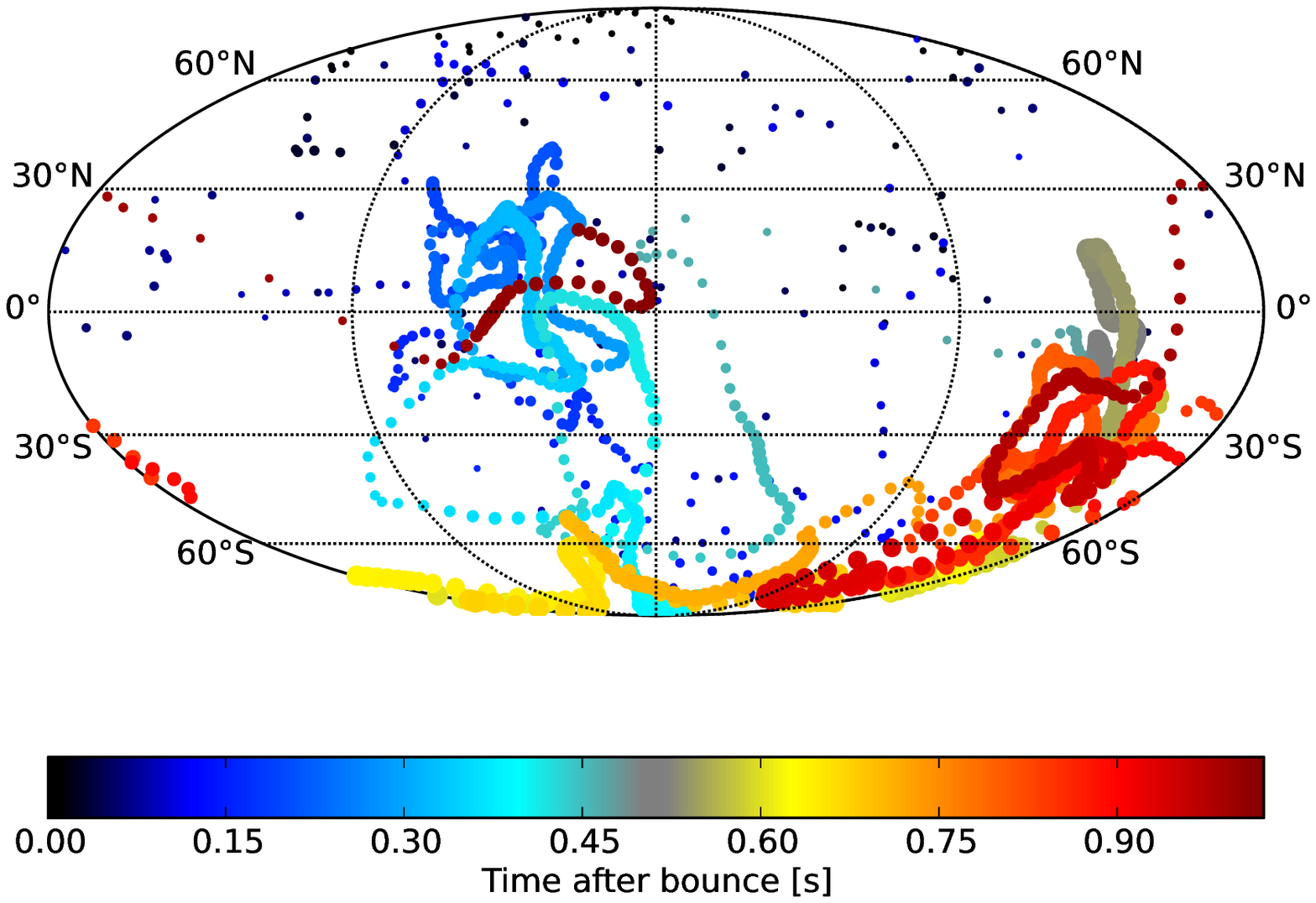}
\null\vskip-0.4in
\includegraphics[height=0.45\textheight]{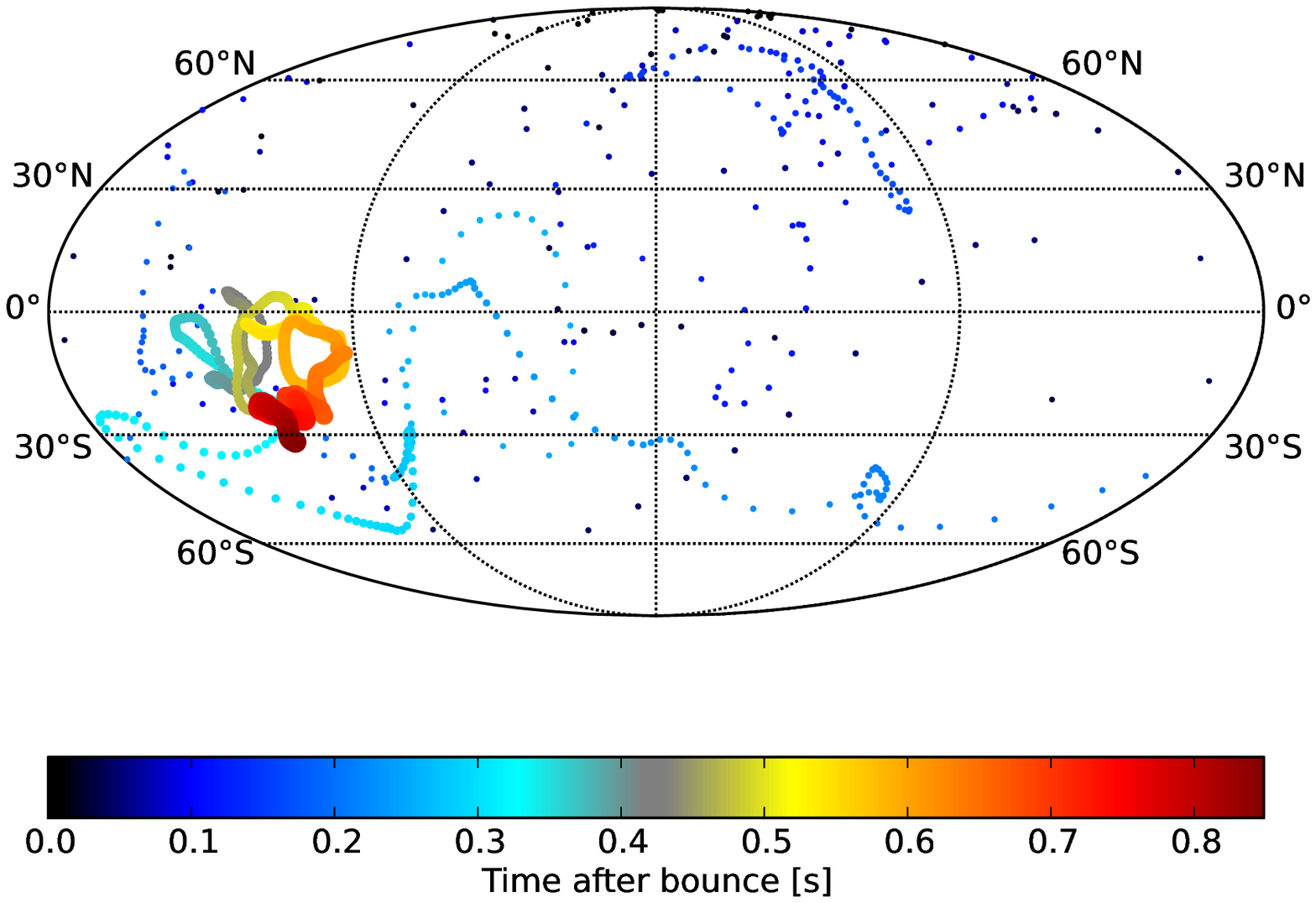}
\end{center}
\caption{Mollweide projection of the $l = 1$ mode in 3D
of its instantaneous vector direction. Two models are shown,
with constant $L_{\nu_{e}}$s of either $2.1\times 10^{52}$ ergs s$^{-1}$ (top)
or $2.2\times 10^{52}$ ergs s$^{-1}$ (bottom). No initial
rotation has been imposed. The dot color indicates time after bounce, the dot size is
proportional to the magnitude of the dipole, and the dot position gives the direction in which
the dipole points. Note that the dot size scale is different on the two panels.
The $m = (0,\pm{1})$ submodes define a principal axis along which
the dipolar mode is instantaneously pointing. Note that the dipole does not oscillate
up and down, but jumps stochastically in the early phases, eventually to find and maintain a
general direction, even as it changes in magnitude. See text for a discussion.}
\label{fig3}
\end{figure}

\clearpage

\begin{figure}
\begin{center}
\includegraphics[height=.4\textheight]{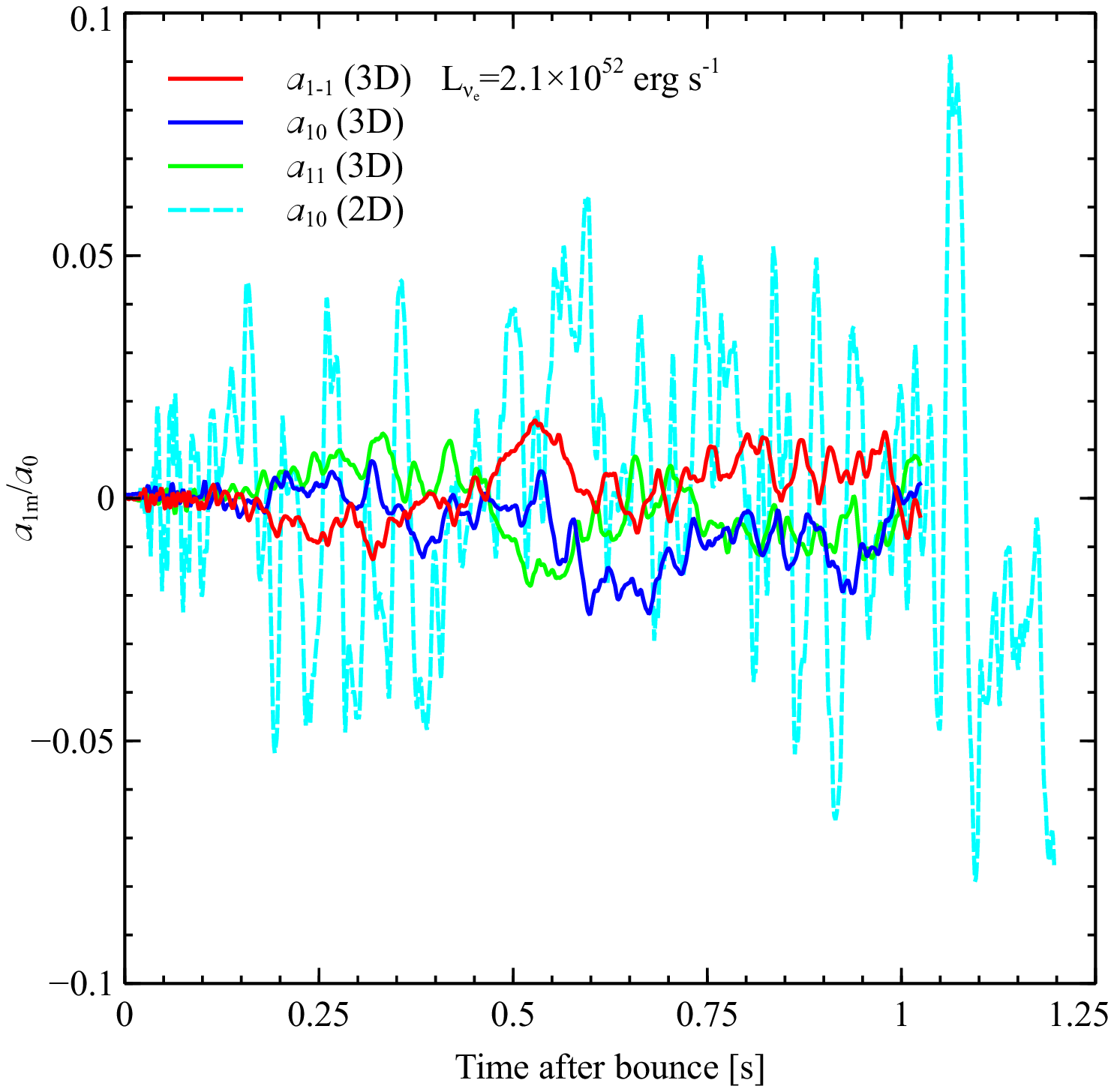}
\includegraphics[height=.4\textheight]{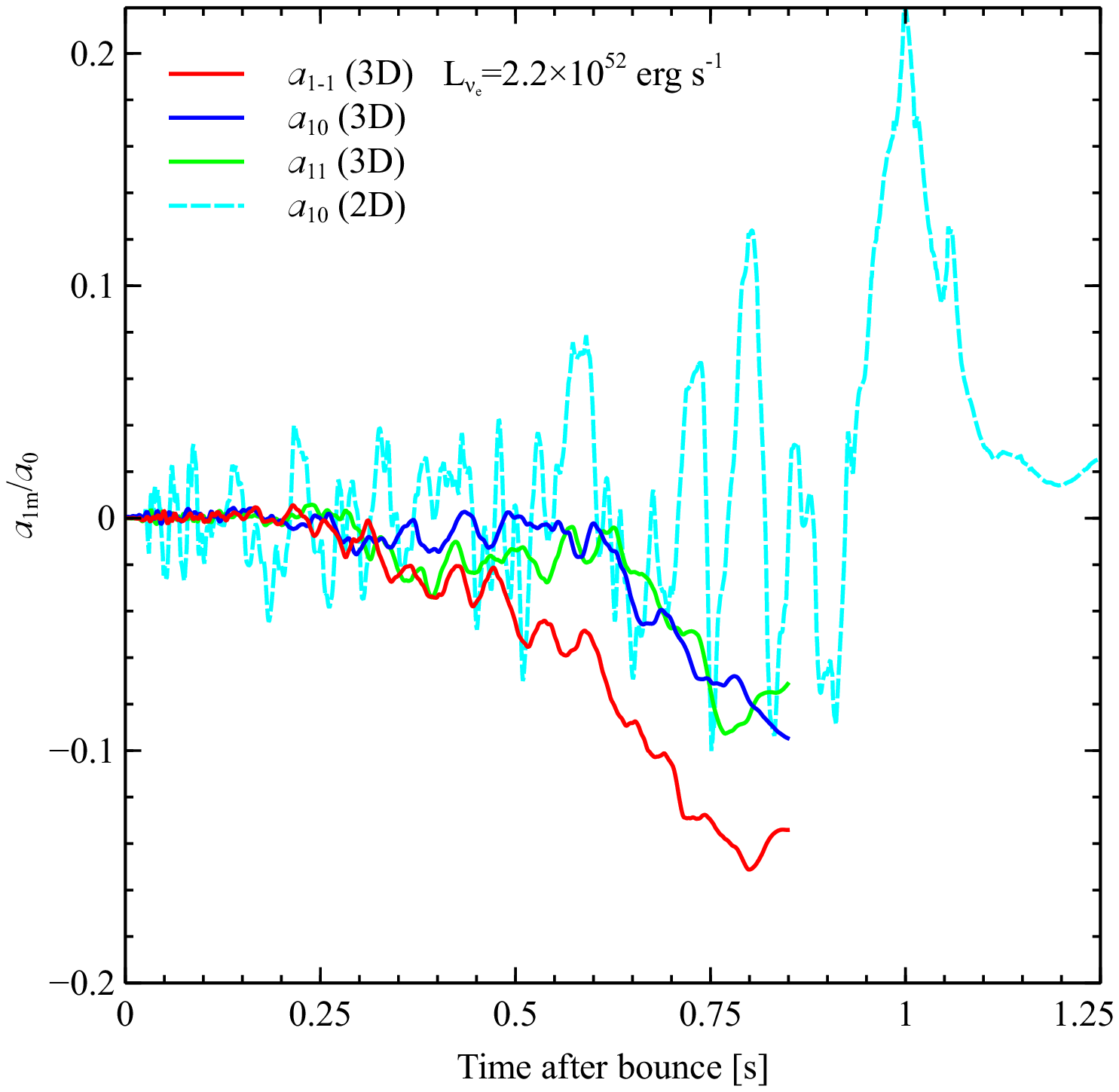}
\end{center}
\caption{The evolution of the individual $m=0,\pm{1}$ harmonic coefficents of 
the dipole ($l=1$) mode of the shock wave surface in 2D (aqua, only $m=0$) and 3D.
Note that the dipolar coefficient in 2D changes sign quasi-periodically with large amplitude, 
while the coefficients in 3D generally do not.  The top panel is for $L_{\nu_{e}} = 2.1\times 10^{52}$ ergs s$^{-1}$
and the bottom panel is for $L_{\nu_{e}} = 2.2\times 10^{52}$ ergs s$^{-1}$. See Figure \ref{fig3} and refer to the text
for a discussion.}
\label{fig4}
\end{figure}

\clearpage

\begin{figure}
\begin{center}
\includegraphics[height=0.25\textheight]{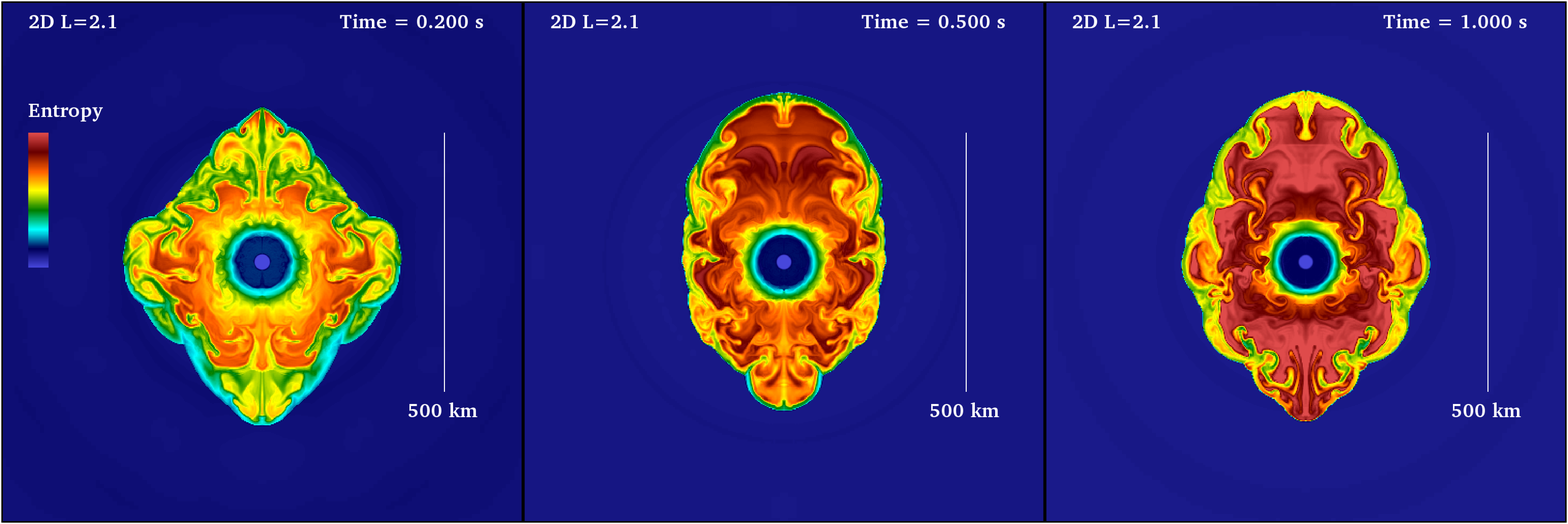}
\includegraphics[height=0.25\textheight]{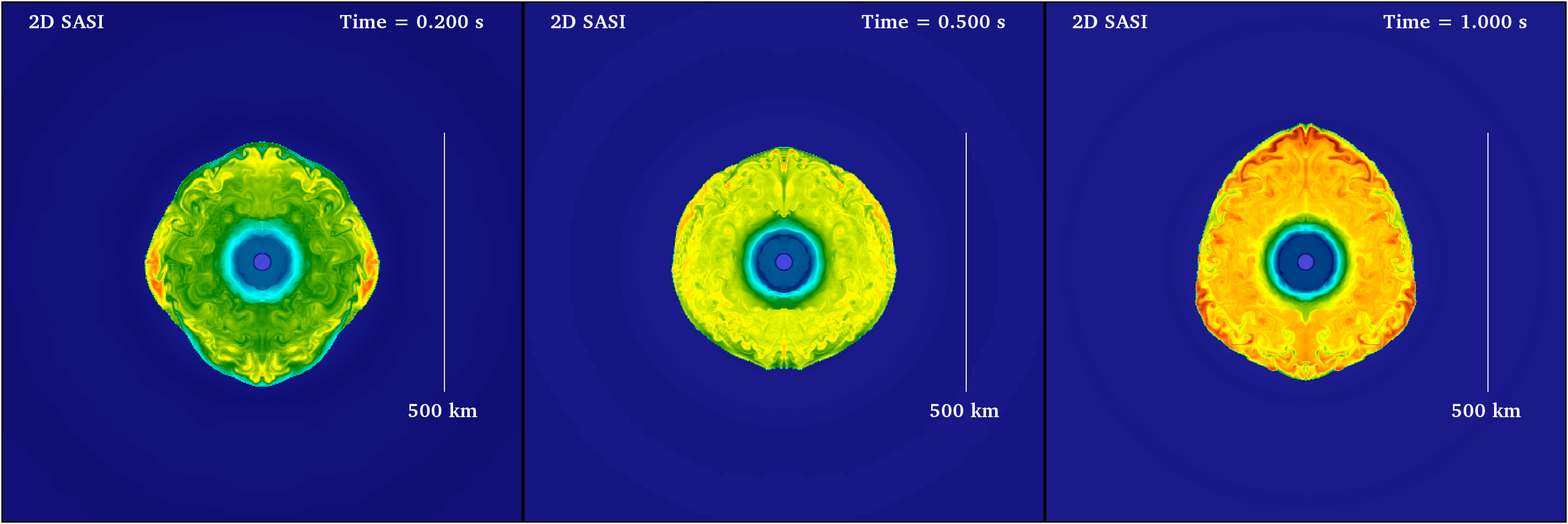}
\end{center}
\caption{A comparison of slices of two 2D models, with (top) and without (bottom) neutrino heating 
(and, hence, for the latter without neutrino-driven convection).  The 2D neutrino-heating 
model employs $L_{\nu_e}$ = $2.1\times 10^{52}$ ergs s$^{-1}$.  Shown are entropy 
maps 200, 500, and 1000 milliseconds after bounce. The color scale ranges from 2.5 (blue) to 
17 (red) $k_b/\textrm{baryon}$. The amplitudes of the shock asymmetries 
depicted are representative of stills throughout the $\sim$1 second of the 
simulations and clearly show how much more quiescent this ``no-neutrino-heating" 
model is.  Note also that this model, for which only the SASI is responsible for 
the 2D motions and entropy generation, has lower entropies in the turbulent region.
This behavior is representative of models without neutrino heating.}
\label{fig5}
\end{figure}

\clearpage

\begin{figure}
\begin{center}
\includegraphics[height=.5\textheight]{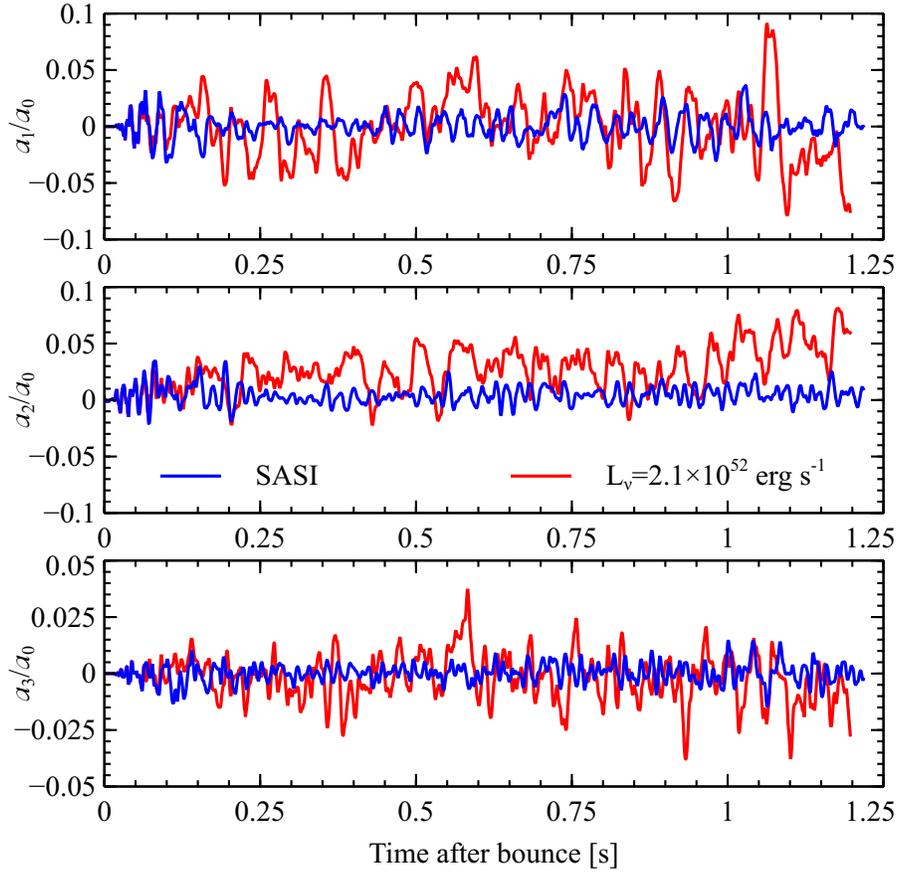}
\end{center}
\caption{The evolution of the $l = 1,2,3$ harmonic coefficients 
(normalized to the $l = 0$ mode) of the stalled shock surface for 
two 2D models, one with (red) neutrino heating ($L_{\nu_e}$ = $2.1\times 10^{52}$ ergs s$^{-1}$)
and one without (blue) neutrino heating, versus time (in seconds).
The plot depicts $\sim$1 second of post-bounce evolution.
The amplitudes of all modes are significantly smaller without neutrino heating and the
associated buoyant convection. See text for a discussion.}
\label{fig6}
\end{figure}

\clearpage

\begin{figure}
\begin{center}
\includegraphics[height=.5\textheight]{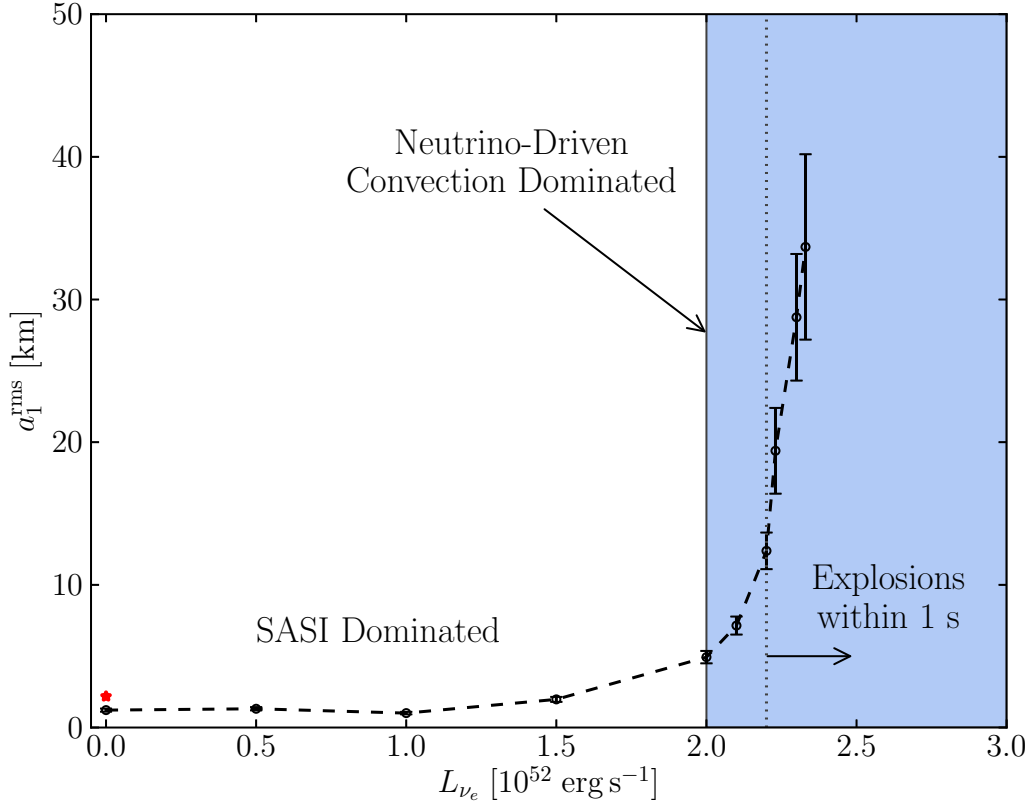}
\end{center}
\caption{Root-mean-square amplitude $\sqrt{\langle a_1^2\rangle}$ (not normalized by 
the mean shock radius) of the dipolar shock deformation as a function of
neutrino luminosity.  For these calculations, the cooling rate is 100\%. 
Below $L_{\nu_e}\sim2\times10^{52}\,{\rm erg\,s}^{-1}$, the amplitudes remain small and
roughly constant even though the average shock radii change by a factor $\sim$two.  For higher luminosities
(shaded region), the amplitudes grow steeply, which we associate with the onset of vigorous neutrino-driven
convection. This is consistent with the dramatic increase in low-frequency power seen in Figure~\ref{fig8}.
Models that explode within one second after bounce are on and to the right of the dotted line.  
In this figure and in Figure \ref{fig8}, we identify the luminosity ranges for which we hypothesize
one can conclude the models are ``SASI-dominated" or ``Neutrino-Driven Convection Dominated." Only data
beyond 250 ms after bounce, but before explosion (if applicable), are included in this analysis.  Note that the
error bars shown are simply $3 \sqrt{\langle a_1^2\rangle/N}$ ($N$ is the number of data points in a given
curve), and are therefore meant only to be suggestive of the relative uncertainty. The red point shows
the results for the zero heating, but reduced (10\%) cooling model.
The dashed line is merely a guide to the eye. Note that a plot of the rms dipolar amplitude 
normalized by the mean shock radius would show the same systematics. (See the text for further discussion.)
}
\label{fig7}
\end{figure}

\clearpage

\begin{figure}
\begin{center}
\includegraphics[height=.4\textheight]{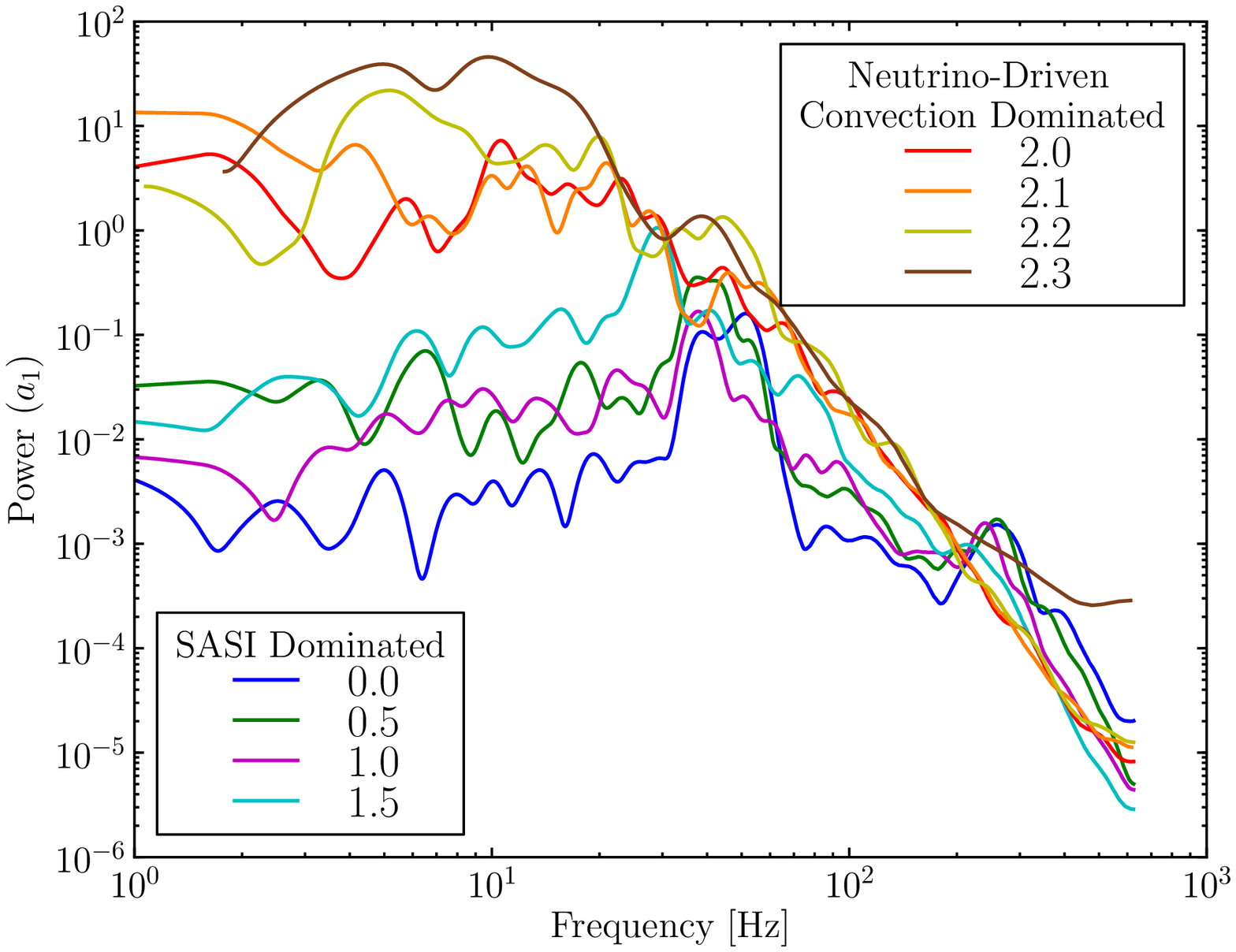}
\includegraphics[height=.4\textheight]{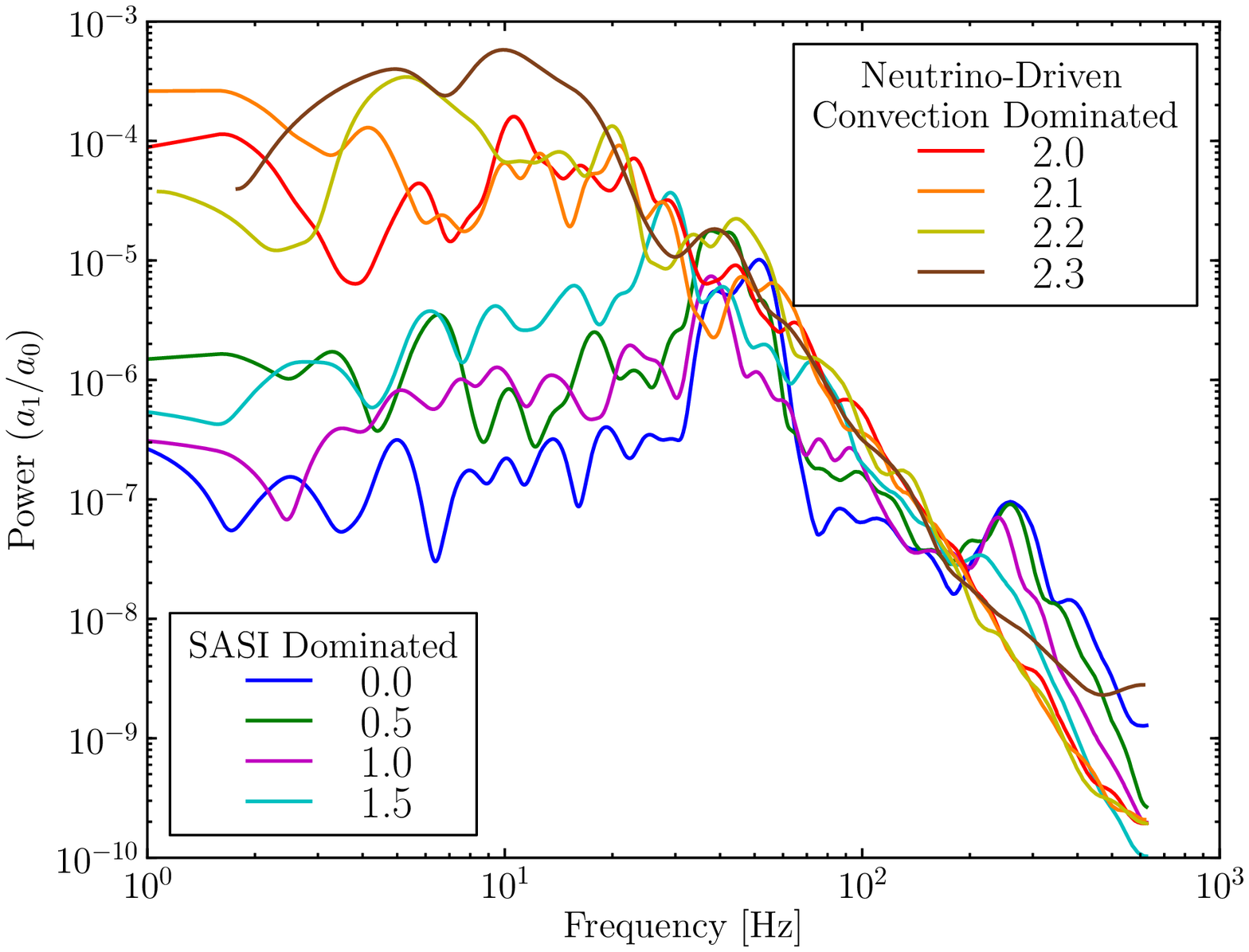}
\end{center}
\caption{Power spectra of the dipolar shock deformation $a_1(t)$ (panel a) and $a_1(t)/a_{0}(t)$ (panel b)
for models with different neutrino luminosities (in units of $10^{52}\,{\rm erg\,s}^{-1}$).  
Models with $L_{\nu_e}\ge2\times10^{52}\,{\rm erg\,s}^{-1}$ show $\sim$100 times more 
low-frequency power than models with lower luminosities.  Note that the excess low-frequency 
power is \emph{not} associated with explosion; models with $L_{\nu_e}\le2.1\times10^{52}\,{\rm
erg\,s}^{-1}$ do not explode within the simulated time and only times before explosion are included in this
analysis for more luminous models.  In all cases, we include only data beyond 250 ms after bounce and smooth
the power spectra for clarity.
}
\label{fig8}
\end{figure}

\clearpage

\begin{figure}
\begin{center}
\includegraphics[height=.5\textheight]{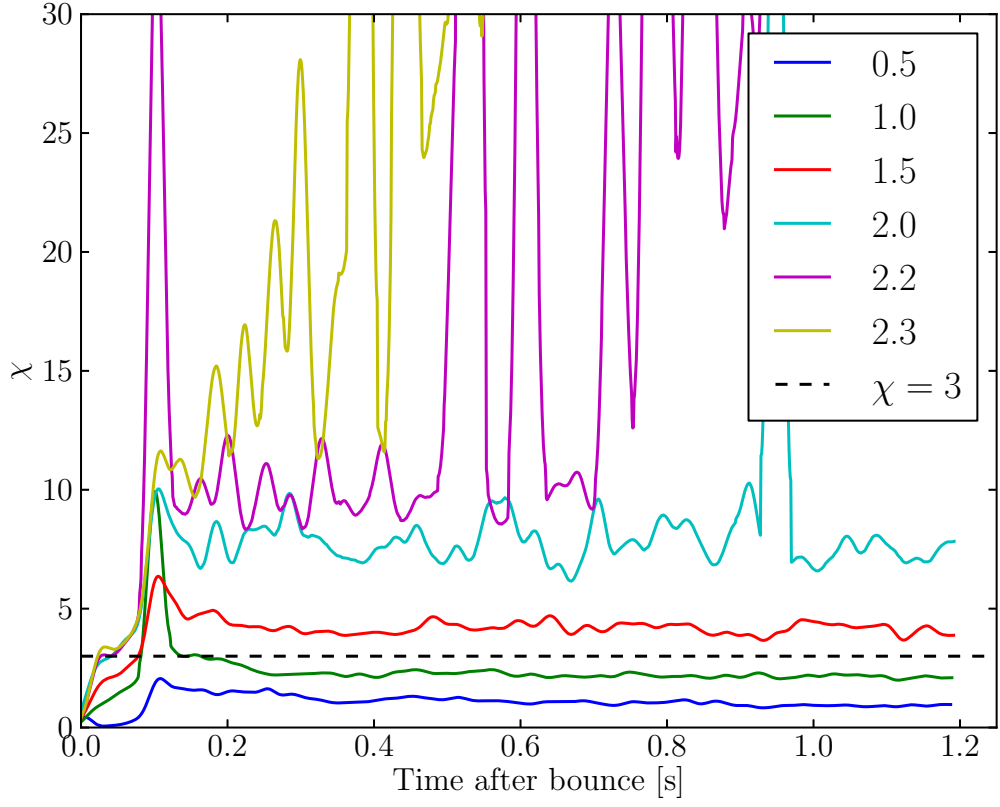}
\end{center}
\caption{Evolution of $\chi$ for models with different neutrino luminosities (in units of $10^{52}\,{\rm erg}\,{\rm s}^{-1}$).  Models
with $L_{\nu_e}\lesssim 1.5\times10^{52}\,{\rm erg}\,{\rm s}^{-1}$ show an evolution of $\chi$ with little temporal fluctuation
and saturation well below the higher luminosity models.  For the higher luminosity models, the vigor of the turbulence
increases with neutrino luminosity and leads to more frequent zero crossings in the radial velocity which produces spikes in $\chi$ as
the integral becomes divergent (see eq. \ref{brunt}).  Note that the curves have been smoothed for clarity. See text for a discussion.
}
\label{fig9}
\end{figure}

\clearpage

\begin{figure}
\begin{center}
\includegraphics[height=.5\textheight]{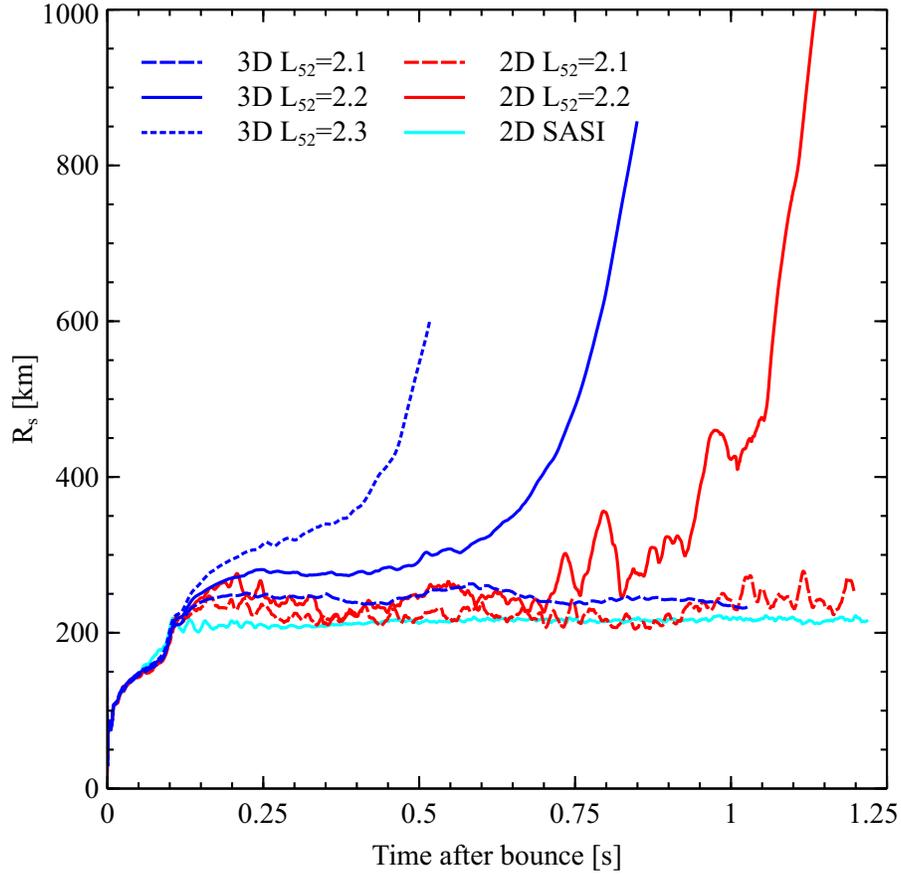}
\end{center}
\caption{The evolution of the average shock radius (R$_s$ = $a_{0}$) in 3D (blue) 
for models with constant driving luminosities ($L_{\nu_e}$) of $2.1\times 10^{52}$ ergs s$^{-1}$,
$2.2\times 10^{52}$ ergs s$^{-1}$, and $2.3\times 10^{52}$ ergs s$^{-1}$ and in 2D 
(red) for models with constant driving luminosities of $2.1\times 10^{52}$ ergs s$^{-1}$
and $2.2\times 10^{52}$ ergs s$^{-1}$.  Higher $L_{\nu_e}$s explode earlier
and the $L_{\nu_e}$ = $2.1\times 10^{52}$ ergs s$^{-1}$ model does not explode at 
all during the times shown. Also portrayed (in aqua) is the corresponding plot 
for the no-neutrino SASI run in 2D with 10\% cooling. See text for a discussion.}
\label{fig10}
\end{figure}

\end{document}